\documentclass[amsmath]{revtex4}

\newcommand{\ssy}[5]{#1,    #2 {\bf #3}, #5 (#4)\rlap{.}}
\newcommand{\evalat}[3]{\left.#1\right|_{#2}^{#3}}%

\begin{document}

\title{Comment on `` Observing a wormhole"}
\author{S. Krasnikov}\email{krasnikov.xxi@gmail.com}
\affiliation{Central Astronomical Observatory at Pulkovo, St.Petersburg, 196140,
 Russia}

\begin{abstract}
In their recent  paper Dai and  Stojkovic discuss an interesting possibility: a star near a wormhole mouth   may gravitationally feel an object located near the other mouth. This means that  a star's trajectory may tell an observer that the star orbits a wormhole mouth and not a black hole. I argue  that within the approximation used in the paper the effect is, in fact, unobservable irrespective of how accurate the measurements are.
\end{abstract}

\maketitle

There is almost consensus that  the massive compact objects in the centers of galaxies  are giant black holes. There is, however,  an alternative point of view, advocated by Kardashev e.~a.~\cite{kar}. According to it,  these objects are mouths of wormholes. It would be important therefore, to find effects distinguishing between these two possibilities, see \cite{foils,WHvsBH}, in particular. One such effect  has been discovered recently \cite{DS} by Dai and  Stojkovic (DS). They noticed that a star orbiting a wormhole mouth may  be affected by  perturbations in the gravitational field  that are produced by an object orbiting the other mouth.

To estimate the effect DS consider a wormhole obtained by gluing together the tubes $r_{1,2}=R$ in a pair of equal portions $r_{1,2}\geq R> r_g $ of Schwarzschild space.
Here $r_{2 ,1}$ are the radial coordinates in, correspondingly,  ``our universe" (the half of the spacetime in which the test star orbits) and the ``other universe", to which the wormhole leads. Now suppose there is an object of a small mass $\mu$ in the ``other universe" at $ r_1 = A>R$ (so, the object is approximated by a light sphere); the radius, $A$, can---quasistatically---vary. The metrics in our and in the other universes differ from that in the case $\mu=0$ by perturbations $\boldsymbol h^{our}$ and $\boldsymbol h^{oth}$ which are assumed to obey the following conditions:
(i) only the components $h_{tt}^{our}$, $h^{our}_{rr}$, $h^{oth}_{tt}$, and $h^{oth}_{rr}$ are non-zero;
   (2)  the perturbations depend only on $r_{1,2}$; 
(3) $h_{\alpha\beta}^{our}(R) = h_{\alpha\beta}^{oth}(R)$ and $\partial_{r_2} h_{\alpha\beta}^{our}\evalat{}{r_2=R}{} = \partial_{r_1} h_{\alpha\beta}^{our}\evalat{}{r_1=R}{}$ \footnote{Note that no justification is given in \cite{DS} for the last condition.}.
Applying  these conditions to the expression for the monopole perturbations borrowed from Ref.~[35] of \cite{DS}, DS infer that
\begin{equation}\label{eq:a}
    a\approx-\mu\frac RA\frac1{r_2^2},
\end{equation}
where the ``additional acceleration"  $a (M,\tau )$ is the difference between the total acceleration $a_{tot} (\tau)$ of a (non-relativistic) test star, and the  acceleration $a_M (\tau)$ experienced by the same star \footnote{``the same star"$\equiv$``a star with the same radial coordinate and the same velocity"}
in the Schwarzschild space with mass $M$
\begin{equation}a(M,\tau )\equiv a_{tot} (\tau )- a_{M}(\tau )\label{eq:*}
\end{equation}
($\tau$ parameterizes  the world line of the star). $a$ serves as an indicator: the spacetime in question is a wormhole, \emph{not} a Schwarzschild black hole, if $a(M)\neq 0$ for all $M$ (and for some $\tau$). Note that the indicator is imperfect: if $a(M) = 0$ for some $M$, both geometries are possible.

Acceleration is measured with very high accuracy. So, it might seem that Eq.~\eqref{eq:a} solves the problem  of the remote detection of supermassive  wormholes. Unfortunately, it does not.
Indeed, ``our universe'' is empty, by construction,  and spherically symmetric, by Eqs.~(30)--(35) in \cite{DS}. Therefore, by Birkhoff's theorem,
the test star moves in a static region of the Schwarzschild space of some mass $M_*$. This is by no means anomalous and is fully consistent with the hypothesis that the object orbited by the star is a mere black hole of mass  $M_*$. Or, formally speaking, $a_{tot} (\tau )= a_{M_*}(\tau )$ and hence, by \eqref{eq:*},
 \begin{equation}\label{a=0}
 a(M_*,\tau )=0 \qquad \forall \tau
 \end{equation}
(from this equation  it follows, in particular, that the right hand sides of Eqs.~(36)--(38) in \cite{DS} are actually zeroes) which means, as mentioned above, that the space may or may not be a wormhole.

\emph{Remark}. To identify the error in DS's argument note that our reasoning fully applies to the region $R<r_1< A_{min}$ of the ``other universe". The region is spherically symmetric and empty, therefore it is static. Thus the  perturbations $\boldsymbol h^{oth}$ are actually zero there. This is in  perfect agreement with  Ref.~[35] in \cite{DS}, to which DS refer in justifying  their Eqs.~(28)--(35) and which, in fact,
reads as follows, see item 10.1: ``Inside the orbit,
the perturbation vanishes."

\bigskip

I am grateful to RFBR for financial support under grant No.~18-02-00461``Rotating black holes as the sources of particles with high energy".

\end{document}